\documentclass[twocolumn]{aastex631}
\usepackage{graphicx}
\usepackage{dcolumn}
\usepackage{bm}
\usepackage{float}

\newcommand{\gr}{$\gamma$-ray}
\newcommand{\grs}{$\gamma$-rays}
\newcommand{\fermi}{{\it Fermi}}

\begin{document}

\title{Possible Neutrino Emission from the Pulsar Wind Nebula G63.7+1.1}

\author[0009-0005-9558-4313]{Shunhao Ji}
\affiliation{Department of Astronomy, School of Physics and Astronomy, Key Laboratory of Astroparticle Physics of Yunnan Province, Yunnan University, Kunming 650091, China; jishunhao@mail.ynu.edu.cn, wangzx20@ynu.edu.cn}

\author[0000-0003-1984-3852]{Zhongxiang Wang}
\affiliation{Department of Astronomy, School of Physics and Astronomy, Key Laboratory of Astroparticle Physics of Yunnan Province, Yunnan University, Kunming 650091, China; jishunhao@mail.ynu.edu.cn, wangzx20@ynu.edu.cn}
\affiliation{Shanghai Astronomical Observatory, Chinese Academy of Sciences, 80 Nandan Road, Shanghai 200030, China}

\author{Dong Zheng}
\affiliation{Department of Astronomy, School of Physics and Astronomy, Key Laboratory of Astroparticle Physics of Yunnan Province, Yunnan University, Kunming 650091, China; jishunhao@mail.ynu.edu.cn, wangzx20@ynu.edu.cn}

\author{Jintao Zheng}
\affiliation{Department of Astronomy, School of Physics and Astronomy, Key Laboratory of Astroparticle Physics of Yunnan Province, Yunnan University, Kunming 650091, China; jishunhao@mail.ynu.edu.cn, wangzx20@ynu.edu.cn}

\begin{abstract}
	We report on our finding of an excess of $54^{+16}_{-15}$ 
	neutrinos at the location of the pulsar wind nebula (PWN) G63.7+1.1.
By analyzing the IceCube track-like neutrino data for a group of 14 PWNe,
	which are selected as the targets because of their reportedly 
	association with molecular clouds, G63.7+1.1 is found to be
	the only one detected
	with neutrino emission and the post-trail significance for 
	the detection is 3.2$\sigma$.
	Previously, this PWN was estimated to have an age of 
$\gtrsim$8\,kyr, contain a candidate pulsar detected in X-rays, and have a
distance of $\sim$6\,kpc. More importantly and related to the PWN's possible
neutrino emission, surrounding molecular materials are seen to interact with 
	the PWN. On the basis of these properties, we examine 
	the proton-proton interactions as the process for the neutrino 
	production. The PWN (or the pulsar) can collectively provide
	sufficient energy to power the required high-energy (HE) protons. 
	This possibly first
	neutrino-emitting case in our Galaxy, with problems or other 
	possibilities 
	to be solved or examined, may reveal to us that PWNe are
	the significant Galactic HE neutrino sources.
\end{abstract}

\keywords{Pulsar Wind Nebulae (2215); Gamma-ray sources (633); Pulsars (1306); Neutrino astronomy (1100)}

\section{Introduction} \label{sec:intro}
The IceCube neutrino observatory at the 
South Pole \citep{Aartsen+17} has been detecting the high-energy (HE), TeV--PeV 
neutrinos that are likely of astrophysical origin since 
2013 \citep{Aartsen+13}.
However, only several astrophysical
association cases have thus far been established. The blazar TXS~0506+056 
and its flaring state in 2017 \citep{txs0506a,txs0506b}, nearby Seyfert 
galaxies NGC~1068 and NGC~4151 (respectively at a significance of 
4.2$\sigma$ and $\sim$3$\sigma$) \citep{ngc1068,Ner+24,Abbasi+25}, 
and the Galactic plane (at a 4.5$\sigma$ significance 
level) \citep{gal}. In addition, a few blazars have also been suggested to emit 
HE neutrinos in their flaring state by taking TXS~0506+056 as a typical
blazar-neutrino association case
(see \citealt{Ji+25} and references therein), and several tidal disruption 
events (TDEs) as well similarly because of the spatial and 
temporal coincidences between the TDEs and neutrino 
events \citep{Stein+21,Reusch+22,jiang+23,van+24,yuan+24,Li+24}.
For the Galactic-plane neutrino emission, although it is likely 
diffusive, it could also arise from unresolved point sources 
\citep{gal}.

Pulsars or pulsar wind nebulae (PWNe) have long been suggested as
possible neutrino emitters 
(see e.g., \citealt{Helfand79, Cheng+90,Bednarek+03}). 
One key question is if nuclei could be accelerated from them.
The likely sites for the acceleration presumably would be the magnetosphere 
or wind of a pulsar, or the termination shock of a PWN.
Possible neutrino-production scenarios of pulsars would be those of
accelerated HE nuclei colliding with matter in a PWN
\citep{Cheng+90, Bednarek+97}, surrounding molecular clouds (MCs) or 
high-density interstellar medium (ISM) \citep{Bednarek02}, with the radiation 
field of a pulsar
\citep{Link+05,Link+06,Jiang+07}, or with supernova ejecta for newly born
pulsars \citep{Fang+12,Fang+16}. 
The hadronic scenarios would also work in PWNe 
\citep{Guetta+03,Amato+03,Bednarek+03,Lemoine+15,Di+17}. In
particular, many PWNe have recently been detected at TeV energies
(see e.g., \citealt{hpwn18}) and the Crab nebula even at PeV energies
\citep{Lhaaso21}. Several searches for neutrino emission in the IceCube 
data from identified TeV PWNe have been 
conducted \citep{Aartsen+13a,ice_3yrs,ice_7yr,ice_stacking,ice_10yrs,Aartsen+20},
while no significant excess was found at the locations of those PWNe. 
In this work, we performed a search from a group of PWNe that are 
associated with MCs, where proton-proton ($pp$) interactions would be more 
likely to occur if HE protons are captured by MCs. 
The PWN targets consist of 14 sources systematically found
in the Galactic longitude $l$ and latitude $b$ ranges of 
$1^\circ < l < 230^\circ$ and  $-5^\circ.5 < b < 5^\circ.5$
based on the Milky 
Way Imaging Scroll Painting (MWISP) CO survey data \citep{Zhou+23}.
Using the publicly 
available neutrino dataset of IceCube, we found a neutrino excess with a 
pre-trail (post-trail) significance of 3.9$\sigma$ (3.2$\sigma$) at the 
location of one of the targets, PWN G63.7+1.1.

This PWN was discovered as a filled-center (FC; or Crab-like) 
supernova remnant (SNR) in
the Galactic plane survey conducted at radio frequency 327\,MHz
\citep{Taylor+92} (see also \citealt{Taylor+96}). 
Further radio and far-infrared observational studies
of it and its surrounding ISM were reported in \citealt{Wallace+97}.
At radio frequencies, G63.7+1.1
has a size of $\sim$8$^\prime$ in diameter, likely interacting directly with 
the ISM, while its kinematic distance $D$ was estimated to
be $D\simeq 3.8\pm$1.5\,kpc. FC SNRs are generally believed to contain a PWN, and thus detailed X-ray observational studies of G63.7+1.1 were
carried out \citep{Matheson+16}. Diffuse X-ray emission at the source 
position, with an irregular 4$^\prime.2 \times 3^\prime.2$ shape, was detected.
Combining the morphological and spectral properties of the nebula 
at X-rays with those at radio frequencies, it was concluded that
G63.7+1.1 is an evolved, $\gtrsim$8\,kyr, PWN \citep{Matheson+16}. 
In addition, 
$D$ was updated to be 5.8$\pm$0.9\,kpc and an X-ray point 
source, 3XMM J194753.4+274357 (or CXO J194753.3+274351), was suggested
as the candidate pulsar that powers G63.7+1.1 \citep{Matheson+16}.
There was also a GeV \gr\ source
detected at the position of G63.7+1.1 
with the Large Area Telescope (LAT) onboard {\it the Fermi Gamma-ray 
Space Telescope (Fermi)}, and this source was listed as a PWN 
in the latest (Data Release 4, DR4) 
{\it Fermi} Gamma-ray LAT (FGL) 14-year source catalog 
(4FGL-DR4) \citep{Ballet+23}.
We analyzed the \fermi\ LAT data for this source, 4FGL J1947.7+2744, 
and found  
that it is more likely the pulsar's \gr\ emission. In this paper we report 
these results. 

\section{Data Analysis and Results}\label{data}
\subsection{IceCube data analysis}\label{neutrino}
IceCube track-like neutrino events, originating in charged-current 
interactions of muon (anti-muon) neutrinos with nucleons, are better suited 
to search for the neutrino emissions produced by point-like sources 
because of their good angular resolution ($\lesssim$ 1$^\circ$).
IceCube has released the all-sky track-like neutrino events from Apr. 2008 
to Jul. 2018 \citep{ice_track}. We only selected the data of seasons 
IC86-I--VII, which were obtained with the full 86-string detectors between May 
2011 and Jul. 2018. To discriminate between signal and background events 
(composed of atmospheric neutrinos and diffuse cosmic neutrino backgrounds), 
we employed an unbinned maximum likelihood method, implemented in {\tt SkyLLH} 
released by IceCube \citep{Wolf+19,IceCube:2021mzg,Bel+24},
to perform the neutrino data analysis \cite{Bra+08}.
In {\tt SkyLLH}, the likelihood function is defined as
\begin{equation}
    \mathcal{L} \left( n_s, \vec{p}_s | D_s \right) = \prod_{i=1}^N \left[ \frac{n_s}{N_t} S_i \left( \vec{p}_s \right) + \left( 1 - \frac{n_s}{N_t} \right) B_i \right],
    \label{eq:likelihood}
\end{equation}
where $n_s$ is the number of signal events in the data sample $D_s$ of 
$N_t$ total 
events. $\vec{p}_s$ represents the set of model parameters of a point-like 
source and contains source position 
$\vec{d}_{\text{src}} = (\alpha_{\text{src}}, \delta_{\text{src}})$ and 
spectral index $\gamma$ of a power-law spectrum in neutrino energy $E_\nu$ 
(i.e., flux $\Phi_{\nu_\mu+\bar\nu_\mu}\propto E_\nu^{-\gamma}$). 
$S_i$ and $B_i$ are the values of the signal and background probability 
density functions (PDFs) for the $i$th data event, respectively, and both 
signal and background PDFs are separated into a spatial and an energy part 
(see the {\tt SkyLLH} document\footnote{\url{https://github.com/icecube/skyllh/blob/master/doc/user_manual.pdf}} for details).

\begin{figure}[!t]
    \includegraphics[width=0.49\textwidth]{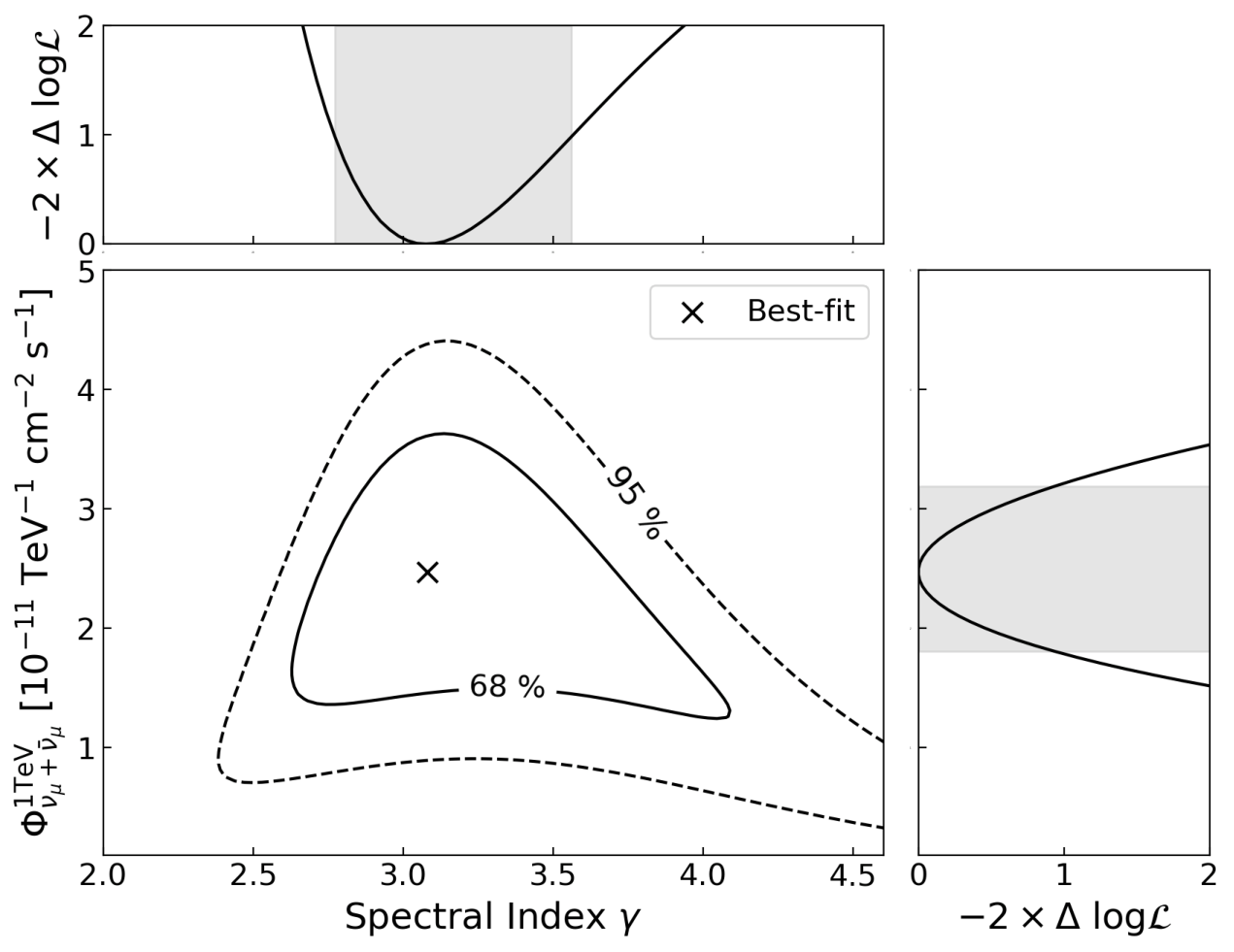}
	\caption{Likelihood scan for the flux parameters of G63.7+1.1. 
	Solid and dashed lines respectively represent 68\% and 95\% CL contours
	derived from using the Wilks' Theorem. The cross is the best-fit value.
	The side panels are the corresponding one-dimensional profile and 
	the gray regions represent the 68\% CL uncertainties. }
    \label{fig:llh}
\end{figure}

The test statistic (TS) is defined as the log-likelihood ratio to test the presence of a signal:
\begin{equation}
    \mathcal{\rm TS} =  -2 \log \frac{\mathcal{L} \left( n_s = 0\right)}{\mathcal{L} \left( n_s, \vec{p}_s\right)}.
\end{equation}
To estimate the significance of a source, we generated the TS distribution 
of background-only data trials for the source. 
Following \citet{ngc1068}, for TS $<$ 5, the p-value (pre-trail) was
estimated directly from the background distribution; for TS $\geq$ 5, 
the truncated gamma function was used to extrapolate the distribution to 
obtain the p-value (see Figure~\ref{fig:back} as an example).

\begin{figure*}
\centering
\includegraphics[height=0.39\textwidth]{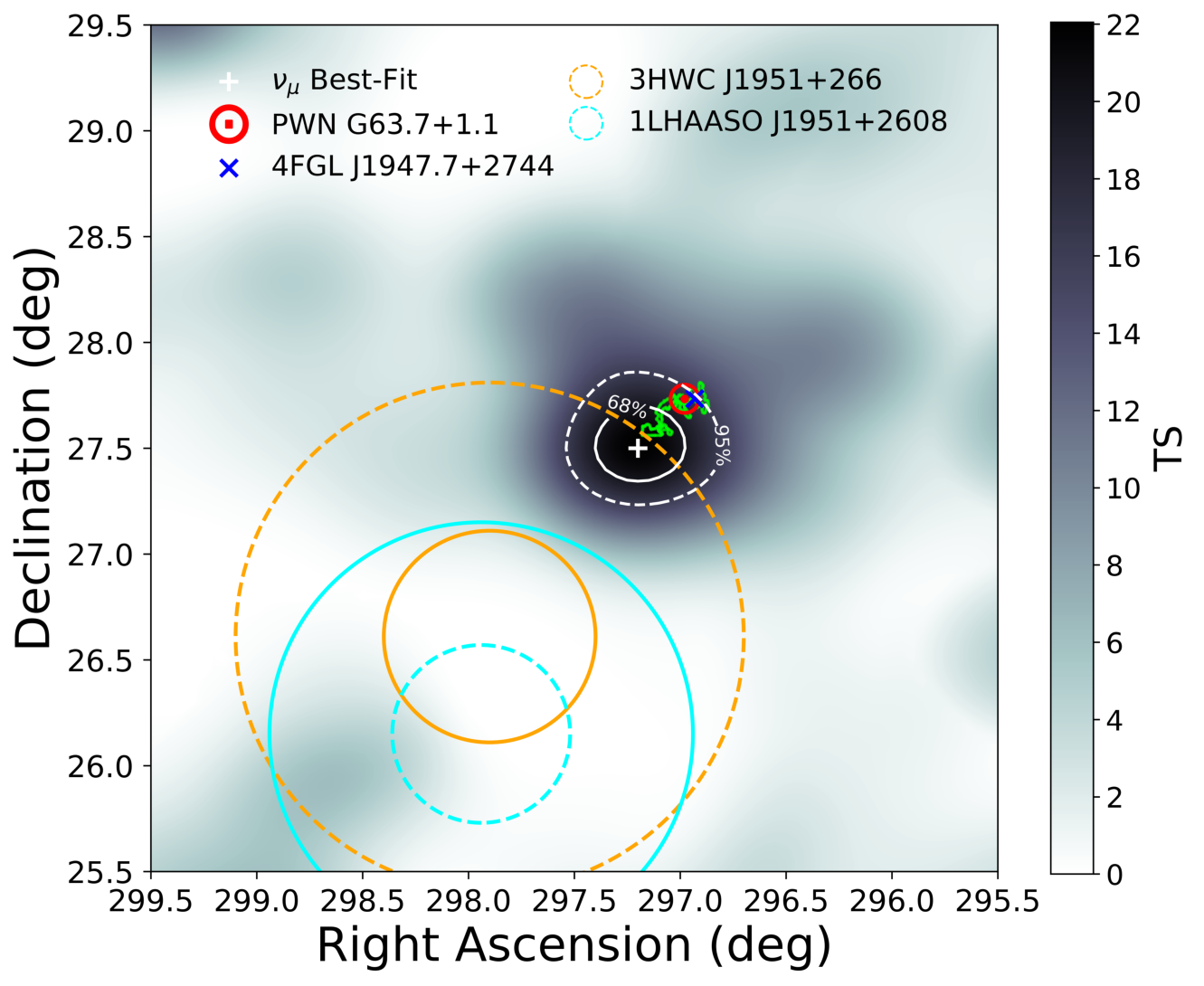}
\includegraphics[height=0.39\textwidth]{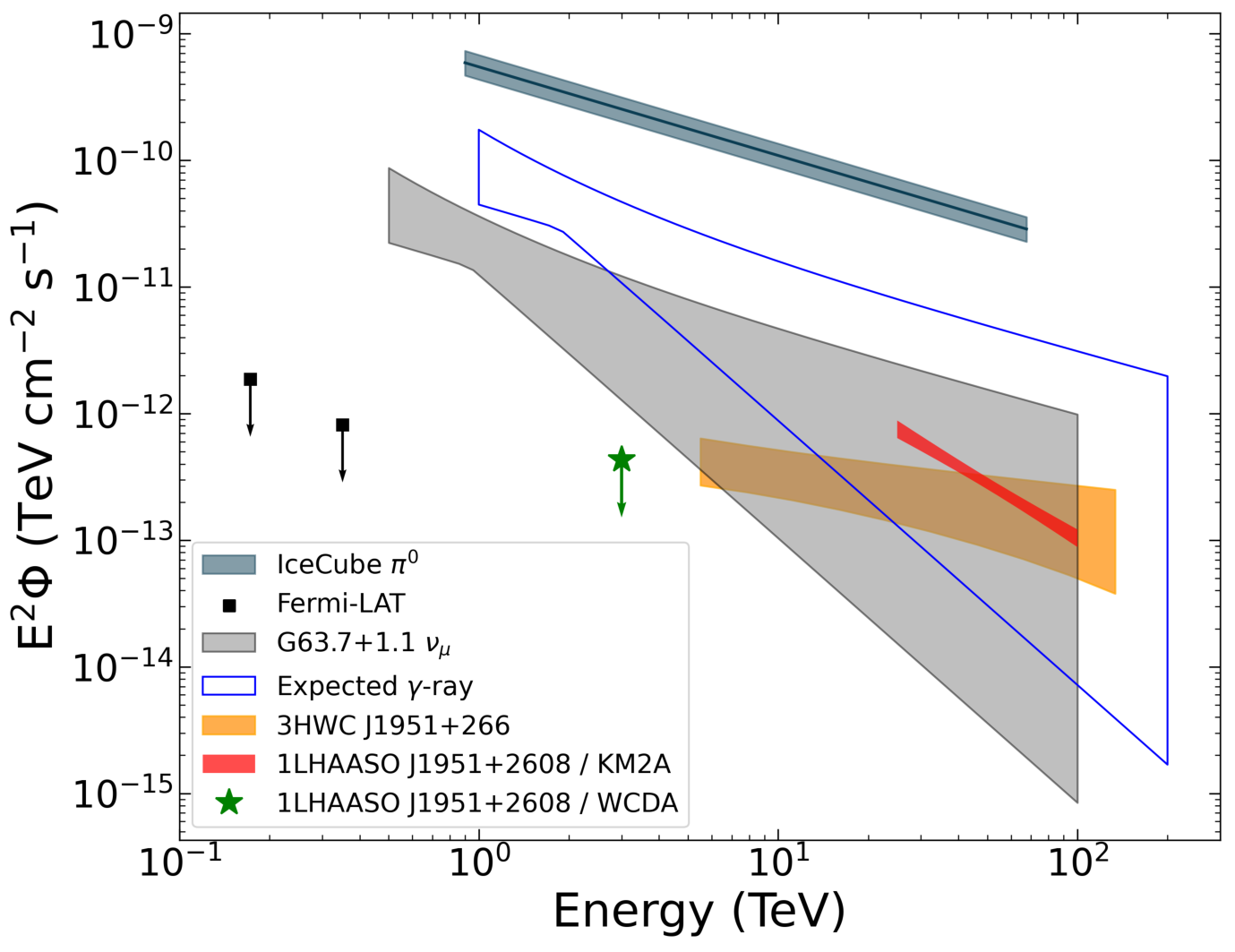}
	\caption{{\it Left:} IceCube neutrino TS map of the G63.7+1.1 region. 
	The hotspot is indicated by the plus sign, with white solid (dashed) 
	contour marking its 68\% (95\%) uncertainty region.
	G63.7+1.1, indicated by a 8$^\prime$ red circle, is within the 
	uncertainty region.
	The \fermi\ source, 4FGL J1947.7+2744 (blue cross), and other TeV 
	sources (yellow and cyan solid and dashed circles) are also marked
	\citep{Albert+20,Cao+24}. 
	The representive profiles of molecular materials around G63.7+1.1,
	reported in \citealt{Wallace+97}, are shown as green lines within
	the uncertainty region of the hotspot.
	{\it Right:} observed neutrino spectrum (gray region) from G63.7+1.1 
	with its 1$\sigma$ uncertainty in 0.5--100\,TeV and 
	expected \gr\ spectrum estimated from neutrino spectrum (blue line). 
	Also shown are the PL model flux (yellow region) of 3HWC~J1951+266 
	\citep{Albert+20} 
	and PL model flux (red region) in 25--100\,TeV of 1LHAASO~J1951+2608
	and its flux upper limit (green star) at 3\,TeV \citep{Cao+24}.
	Black squares 
	are the flux upper limits on 4FGL J1947.7+2744. The neutrino flux of 
	the Galactic plane using $\pi^0$ model \citep{gal} is plotted 
	(dark blue region) for comparison.
\label{fig:tsmap}}
\end{figure*}

We performed the likelihood analysis to the IceCube data for each of the 14 
PWN sources, and found that the location of G63.7+1.1 had
a TS value of 17.3 and a pre-trail p-value of $5.3\times10^{-5}$ (3.9$\sigma$),
the most and only significant one among our targets. 
In Figure \ref{fig:back}, the background TS distribution 
at the location of G63.7+1.1 is shown. The information for the 
PWNe and neutrino likelihood analysis results are
provided in Table~\ref{tab:tgt}. 
We evaluated the post-trial p-value from 
$1 - (1-p_{pre})^N$, where $p_{pre}$ is pre-trail p-value and $N$ is 
the number of sources ($N=14$, or the trial factor).
After considering the trial correction, the p-value was 
$7.4\times10^{-4}$ (3.2$\sigma$).  The best-fit parameters we obtained 
are $n_s=54^{+16}_{-15}$ and $\gamma=3.1^{+0.5}_{-0.3}$, 
and flux at 1\,TeV $\Phi^{1\rm TeV}_{\nu_\mu+\bar\nu_\mu}=2.5^{+0.7}_{-0.7}\times10^{-11}~\rm TeV^{-1}cm^{-2}s^{-1}$. 
All uncertainties, derived from the likelihood scan (Figure \ref{fig:llh}),
are at a 68\% confidence level (CL).

We further obtained the TS map by scanning a $4^\circ\times4^\circ$ region 
with a binsize of $0^\circ.05$ around G63.7+1.1 (left panel of 
Figure \ref{fig:tsmap}).  The hottest location in the map is at 
R.A. = $297^\circ.2$, Decl. = $27^\circ.5$ (J2000.0),
with a TS value of 22.0 and a pretrial p-value 
of $4.2\times10^{-6}$ (4.5$\sigma$).  We estimated the 68\% and 95\% confidence
regions for the hotspot using the Wilks' Theorem \citep{Wilks} with two degrees 
of freedom \citep{ice_7yr}.
As indicated in Figure \ref{fig:tsmap}, G63.7+1.1 is within the 95\% 
confidence 
region of the hotspot and has a $\sim$$0^\circ.3$ offset from the hottest 
location (note that such a small offset has been seen in IceCube data analysis 
using the 10-year dataset; see e.g., \citealt{ice_10yrs,Ner+24}). 
We checked the SIMBAD database\footnote{\url{https://simbad.cds.unistra.fr/simbad/}} 
within the 95\% confidence region of the hotspot, and no other potential 
HE sources
such as active galaxies were found. We noted that this neutrino hotspot was 
also reported in the Northern sky scan \citep{ngc1068} and Galactic plane 
scan \citep{Abbasi+23,Li+25} using the IceCube data. Interestingly, this 
hotspot is the second most significant location in the Northern sky, with 
the first being NGC 1068 \citep{ngc1068}, and the most significant location in 
the Galactic plane \citep{Li+25}. Our scan results are consistent with those 
in \citet{ngc1068} and \citet{Li+25}.

\subsection{LAT Data analysis for 4FGL J1947.7+2744}
\label{sec:fermi}
We analyzed the \fermi\ LAT data
for 4FGL J1947.7+2744 because of its positional coincidence with
G63.7+1.1 and the neutrino hotspot (Figure~\ref{fig:tsmap}).
While this \fermi\ source is listed as a PWN in 4FGL-DR4 \citep{Ballet+23},
we found that its emission in 0.3--500\,GeV can be well described with a 
typical model for pulsars. Also no extension was detected.
By mainly considering X-ray emissions of the PWN and pulsar 
candidate \citep{Matheson+16} and the upper limit on the TeV emission of 
the PWN, we suggested that the \gr\ source, rather than being the PWN,
is more likely the pulsar. 
The detailed data-analysis processes are described in 
Appendix~\ref{sec:fermiana},
and the arguments for 4FGL J1947.7+2744's pulsar origin are presented
in Appendix \ref{sec:psr}.

\begin{table*}[!ht]
    \centering
	\caption{List of 14 PWN-MC targets}
    \begin{tabular}{lccccccccc}
        \hline
        PWN & TeV Counterpart & Age & RA & DEC & TS & $n_s$ & $\gamma$ & $-\log_{10} p_{pre}$ & $\Phi^{90\%}_{\nu_\mu+\bar\nu_\mu}$$^*$\\
	    & & (kyr) & (deg) & (deg) & & & & &  \\
        \hline
	    Crab & HESS J0534+220 & 0.971$^{\dagger}$ & 83.63 & 22.02 & 0.6 & 7.9 & 5.00 & 0.45 (0.4$\sigma$) & 10.6 \\
        Geminga & MGRO J0632+17 & 338.8 & 98.48 & 17.77 & 0.1 & 2.7 & 3.41 & 0.27 (0.0$\sigma$)& 5.5 \\
        Eel & HESS J1826$-$130 & 14.4 & 276.53 & $-$13.0 & 0.0 & 0.0 & 4.00 & 0.00 (0.0$\sigma$) & 12.5 \\
        G18.00$-$0.69 & HESS J1825$-$137 & 21.4 & 276.55 & $-$13.58 & 0.0 & 0.0 & 5.00 & 0.00 (0.0$\sigma$) & 8.7 \\
        G21.88$-$0.10 & HESS J1831$-$098 & 128.0 & 277.86 & $-$9.87 & 0.0 & 0.0 & 3.34 & 0.00 (0.0$\sigma$) & 6.1  \\
        G23.5+0.1 & ... & 147.9 & 278.42 & $-$8.46 & 0.4 & 1.5 & 2.71 & 0.35 (0.1$\sigma$) & 12.2\\
        G25.24$-$0.19 & HESS J1837$-$069 & 23.0 & 279.51 & $-$6.93 & 0.0 & 0.0 & 1.00 & 0.00 (0.0$\sigma$) & 3.5 \\
        G32.64+0.53 & HESS J1849$-$000 & 42.9 & 282.26 & $-$0.02 & 0.0 & 0.0 & 5.00 & 0.00 (0.0$\sigma$) & 3.2 \\
        G36.01+0.06 & MAGIC J1857.2+0263 & 20.6 & 284.21 & 2.76 & 0.0 & 0.0 & 4.34 & 0.00 (0.0$\sigma$) & 1.9 \\
        G47.38$-$3.88 & ... & 3090.3 & 293.06 & 10.99 & 0.2 & 3.9 & 3.09 & 0.33 (0.1$\sigma$) & 3.7 \\
	    \textbf{G63.7+1.1}$^{\ddagger}$ & \textbf{1LHAASO J1951+2608} & \textbf{27.0} & \textbf{296.98} & \textbf{27.73} & \textbf{17.3} & \textbf{54.1} & \textbf{3.08} & \textbf{4.28 (3.9$\sigma$)} & \textbf{32.6} \\
         & \textbf{(3HWC~J1951+266)?} &  &  &  &  &  &  &  &  \\
        CTB 87 & VER J2016+371 & 4.0--28.0 & 304.04 & 37.19 & 0.5 & 7.2 & 3.75 & 0.47 (0.4$\sigma$) & 13.2 \\
        G75.23+0.12 & MGRO J2019+37 & 17.0 & 305.27 & 36.85 & 0.0 & 0.0 & 5.00 & 0.00 (0.0$\sigma$) & 7.5 \\
        G80.22+1.02 & TeV J2032+4130 & 110.0 & 308.05 & 41.46 & 4.6 & 22.8 & 3.33 & 1.38 (1.7$\sigma$) & 20.3 \\
        \hline
    \end{tabular}
    \tablecomments{The 14 PWN-MC sources reported in \citealt{Zhou+23}, for which the TeV counterparts and pulsars' characteristic ages (except the Crab pulsar) are  from SNRcat\footnote{\url{http://snrcat.physics.umanitoba.ca}} \citep{snrcat}. The characteristic age of the putative pulsar in G63.7+1.1 is from  \citealt{Matheson+16}. 
    Among the targets, 11 of them have TeV \gr\ counterparts and G63.7+1.1 is potentially associated with two TeV sources (proposed in this work). $^*$ The 90\% CL upper limits $\Phi^{90\%}_{\nu_\mu+\bar\nu_\mu}$ are at 1 TeV in units of $10^{-13}$ TeV cm$^{-2}$ s$^{-1}$, obtained by assuming the spectral index $\gamma=2$. $^{\dagger}$ This is the supernova age instead. $^{\ddagger}$ G63.7+1.1 is highlighted in bold because of its highest significance.}
    \label{tab:tgt}
\end{table*}

\section{Discussion}\label{dis}
We have conducted a search for neutrino emissions from MC-associated PWNe
in the IceCube track-like neutrino data.
Among 14 of the PWN targets, which were from a CO survey of nearly half of 
the Galactic plane,  we have found neutrino excess at
G63.7+1.1. The finding has a post-trail significance of 3.2$\sigma$.
Thus, this PWN could be the first HE neutrino source found in our Galaxy,
potentially providing a piece of smoking-gun evidence for
hadronic acceleration in PWNe.

As introduced above, pulsars and PWNe have been widely discussed as possible
sources
emitting HE protons. For a PWN, $pp$ interactions is the more likely 
mechanism for neutrino production \citep{Guetta+03,Amato+03}. The HE
protons would collide with target protons inside a PWN or be captured by 
surrounding MCs to generate charged and neutral pions, which subsequently 
decay producing neutrinos and $\gamma$-rays, respectively. 
Comparing G63.7+1.1 to other PWNe, this relatively old PWN is an
unexpected neutrino source \citep{Bednarek+03}. However, when a PWN is surrounded
with high-density ISM, the hadronic emissions could be largely enhanced 
(see e.g. MSH15-52 discussed in \citealt{Bednarek+03}). 
Also some MCs can be located away from a PWN with  
significant distances, delayed hadronic emissions would be expected
with extra time needed for diffusion of protons \citep{Gabici+07}. 
The CO observations revealed that dense molecular materials or MCs exist
at the location of G63.7+1.1 and around it \citep{Wallace+97,Matheson+16,Zhou+23},
which could provide abundant target protons for $pp$ interactions and may be 
responsible for the observed neutrinos from the source. 
Interestingly, the molecular materials reported by \citealt{Wallace+97} 
extends towards the southeast of the PWN, actually matching the hottest 
neutrino location (see Figure \ref{fig:tsmap}). 
Although the small offset of the hottest neutrino 
location from G63.7+1.1 could be due to some systematic 
uncertainties in the data analysis, which are suggested by
similar positional offsets seen in previous IceCube data-analysis results
(see e.g., \citealt{ice_10yrs,Ner+24}), the offset could also be
real, caused by the offsetting $pp$ interactions. 
The hottest position is $\sim0^\circ.3$, or $\sim$31 pc away from G63.7+1.1
(at $D \sim$6\,kpc).
The diffusion coefficient of protons
is given as 
$D_e \sim 3.2\times10^{28}(\chi/0.01)(E_p/1\ \rm PeV)^{0.5}(B/3\ \rm \mu G)^{-0.5}$ cm$^2$~s$^{-1}$ \citep{Gabici+09}, 
here for proton energy $E_p$ and interstellar magnetic field $B$, we assume 
the reduction factor $\chi \sim 0.01$ \citep{Fujita+09}. The diffusion length 
is given as $r_{\rm diff}=2\sqrt{D_e t}$ \citep{Fujita+09}, where $t$ is the 
diffusion time. Given the estimated age of the PWN, $t$ should be 
$\lesssim$8 kyr, and $r_{\rm diff} \lesssim 58~\rm pc$.
It is thus possible for HE protons from G63.7+1.1 to traverse the molecular 
materials in the southeast and interact with them. 

Given the long lifetime of protons in MCs ($>10^5$ yr) \citep{Gabici+09},
the observed neutrinos should be produced from the continuous proton injection
from the PWN. For the spin-down energy $\dot{E}$ of the pulsar 
($\dot{E}\sim 2.1\times 10^{36}$\,erg\,s$^{-1}$; \citealt{Matheson+16}), 
the time-averaged $\bar{\dot{E}}$ can be estimated as
\begin{equation}
	\bar{\dot{E}}=[\int_{0}^{T} \dot{E}_0(1+\frac{t}{\tau_0})^{-\frac{n+1}{n-1}}dt]/T \sim 5.2\times10^{37}\rm erg~s^{-1},
\label{edot}
\end{equation}
where we assume $T=6~\rm kyr$ 
(because 31\,pc needs $\sim$2\,kyr diffusion time),
the initial spin-down timescale $\tau_0 \sim 500$\,yr, and $n=3$ \citep{hpwn18}.
$\dot{E}_0$ is the initial spin-down energy, estimated from 
$\dot{E}$ at present time. The neutrino 
luminosity $L_{\nu_\mu}$ we obtained at $D\sim 6$\,kpc 
was $3.3\times10^{35}$\,erg~s$^{-1}$ 
in 0.5--100\,TeV. The neutrino radiation efficiency 
$L_{\nu_\mu}/\bar{\dot{E}}$ would be $\sim$ $6\times10^{-3}$. 
It is hard to estimate a reliable proton luminosity. 
The differential proton luminosity in $pp$ interactions is given as 
\citep{Murase+16}
\begin{equation}
\epsilon_p L_{\epsilon_p}\approx\frac{6}{{\rm min}[1,~f_{pp}]}\epsilon_{\nu_\mu} L_{\epsilon_{\nu_\mu}},
\label{Lp}
\end{equation}
where energy $\epsilon_p\simeq20\epsilon_\nu$ 
and $\epsilon L_\epsilon=\epsilon^2 \Phi4\pi D^2$,
and $f_{pp}$ is the $pp$ optical depth. 
In a cloud, $f_{pp}$ can be estimated as \citep{Murase+20}
\begin{equation}
f_{pp}\approx t_{\rm esc}/t_{pp}\approx \frac{L_c^2n_p \kappa_{pp} \sigma_{pp}c}{6D_e(\epsilon_p)},
\label{fpp}
\end{equation}
where $\kappa_{pp}\sim0.5$ and $\sigma_{pp}\sim4\times10^{-26}~\rm cm^2$ are 
the proton inelasticity and $pp$ cross section, respectively, 
and $n_p$ is the proton density in a cloud. 
We consider the escape time $t_{\rm esc}$ for protons in the cloud as $t_{\rm esc}\sim t_{\rm diff}\sim L_c^2/6D_e(\epsilon_p)$ \citep{Gabici+09}, here $L_c$ is the size of the could and $D_e(\epsilon_p)$ is the diffusion coefficient for protons with energy of $\epsilon_p$. We assume
a proton density 500 cm$^{-3}$ in a cloud, a cloud size 
of 30 pc, and a magnetic field strength of 60 $\mu\rm G$ \citep{Fujita+09}.
From the assumptions, $f_{pp} \sim 0.04$--0.6, depending on $\epsilon_p$,
and the estimated proton luminosity at $D\sim 6$\,kpc is 
$\sim 6\times10^{36}~\rm erg~s^{-1}$ in 10\,TeV--2\,PeV. The value suggests
that $\sim12\%$ of $\bar{\dot{E}}$ is used for the production of HE protons.

In the hadronic scenario, the relative differential fluxes between pionic \grs\ and all-flavor neutrinos at energies 
$E_{\gamma}\simeq2E_\nu$ are related as \citep{Ahlers+14}
\begin{equation}
E_\gamma^2 {dN_\gamma\over dE_\gamma} \simeq e^{-\frac{D}{\lambda_{\gamma\gamma}}}\frac{4}{K} \frac{1}{3}\sum_{\nu_\alpha} E_\nu^2 {dN_{\nu_\alpha}\over dE_\nu},
\label{flux}
\end{equation}
where $e^{-\frac{D}{\lambda_{\gamma\gamma}}}$ represents the $\gamma$-ray 
absorption by the cosmic microwave background, which is generally negligible 
for a Galactic source \citep{Ahlers+14}. $K$ is the ratio of charged to neutral 
pions and $K \simeq 2$ in $pp$ interactions. 
Assuming full mixing and given the observed muon neutrino spectrum of 
G63.7+1.1, we estimate the expected \gr\ spectrum from Eq. \ref{flux} and 
show it in the right panel of Figure \ref{fig:tsmap}. The \gr\ emission
should be detectable with the current TeV facilities. However, no TeV 
sources were reported at the position of G63.7+1.1.
We note that there are two TeV sources, 3HWC J1951+266 \citep{Albert+20}
and 1LHAASO J1951+2608 \citep{Cao+24}, located close to the neutrino hotspot 
and at the southeast of G63.7+1.1
(Figure~\ref{fig:tsmap}; see also \citealt{Abbasi+23,Li+25}).
The first (with TS $\simeq 35.6$)
has an extension of 0$^\circ.$5 with a positional uncertainty of 
$\sim$1$^\circ.$2 (at a 2$\sigma$ CL),
and the second, only detected in 25--100\,TeV with TS $\simeq 100$,
has an extension of 1$^\circ$\ with a positional uncertainty of 0$^\circ.$42 
(at a 95\% CL). Due to the positional coincidence,
the two TeV sources were marked in association in \citealt{Cao+24}, while
the second one suffered from significant Galactic diffuse emission 
(GDE), which could affect its fitted location and extension. 
Previously, 3HWC J1951+266 was reported to be in potential associate with 
the neutrino hotspot at a significance of 2.6$\sigma$ with a neutrino 
extension of 1$^\circ$.7 (note G63.7+1.1 is also within the extension region; 
\citealt{Abbasi+23}).

We show the spectra of these two TeV sources in the right panel of 
Figure \ref{fig:tsmap} to compare them with the expected \gr\ 
spectrum estimated above.  As can be seen, they are compatible with each other, 
in particular 1LHAASO J1951+2608, which has a spectral index closely matching
that of the expected \gr\ spectrum. We thus suspect that the two TeV 
sources might be the hadronic TeV counterparts to the neutrino hotspot (and to
G63.7+1.1).  Assuming they are, 
the lower limits on the hadronic fractions in their emissions can be estimated.
Within the 1$\sigma$ uncertainty of the neutrino flux, the fractions would 
be $\geq 16\%$ and $\geq 11\%$ at $E_\gamma=50$ TeV for 3HWC J1951+266 and
1LHAASO J1951+2608, respectively.
The values are compatible with the hadronic fraction constrained from the 
Galactic TeV source sample \citep{vec+23}. 

There is one notable problem, however, in the above scenario we have examined:
the expected \gr\ flux is much higher than the observed one at $\sim$TeV 
energies given the non-detection in 1--25 TeV (see Figure \ref{fig:tsmap}). 
This problem may be similar to the general mismatch between $\gamma$-rays and
neutrinos \citep{Murase+16}, which points to the existence of a population of
hidden cosmic-ray accelerators. 
Detailed multi-band studies of G63.7+1.1 are required to probe and determine
properties of its components,
in particular the pulsar. For example, the possible SNR shell 
was mentioned in \citet{Matheson+16}, which might hint on
the consideration of adding the SNR's contribution to the observed
neutrino emission.  Another possibility is the $p\gamma$ scenario for 
pulsars \citep{Link+05,Link+06}; 
the optical depth of $\gamma\gamma \to e^+e^-$ pair productions could be
large near the surface of the pulsar in G63.7+1.1 
($\tau_{\gamma\gamma}\simeq10^3\tau_{p\gamma}$; see e.g., \citealt{Murase+16,Fang+24}), which would make G63.7+1.1 appear like
a `dark' neutrino source. Also, there is a possibility that G63.7+1.1 
may not be the neutrino source. Therefore, as more TeV data are being 
collected, more significant detection 
of the nearby TeV sources could further constrain or confirm their positions 
and extensions, helping clarify the picture at the region.

Finally, G63.7+1.1's neutrino flux is only at $\sim$ 3\% level of that of
the diffuse neutrino
emission in the Galactic plane (see Figure \ref{fig:tsmap}). This PWN
case, if it is confirmed, would suggest many other PWNe as the
Galactic neutrino sources. Among our
target PWNe, about half are in the Southern sky, to which the IceCube 
observations are not at its optimal sensitivity. Near-future neutrino detectors 
located at the Northern hemisphere,
e.g., KM3NeT \citep{km3net} and Baikal-GVD \citep{baikal}, are suited to search
for sources in the Galactic plane and thus could confirm our result by detecting
more neutrino PWNe.

\begin{acknowledgments}
	We thank the referee for the comments that helped reform the
	manuscript. We also thank C. Bellenghi and W. Li for their assistance with our neutrino data analysis. This research is supported by the Basic Research Program of Yunnan Province 
(No. 202201AS070005), the National Natural Science Foundation of China 
(12273033), and the Original Innovation Program of the Chinese Academy of 
Sciences (E085021002).
\end{acknowledgments}

\bibliographystyle{aasjournal}
\bibliography{main_apj}

\begin{thebibliography}{}
\expandafter\ifx\csname natexlab\endcsname\relax\def\natexlab#1{#1}\fi
\providecommand{\url}[1]{\href{#1}{#1}}
\providecommand{\dodoi}[1]{doi:~\href{http://doi.org/#1}{\nolinkurl{#1}}}
\providecommand{\doeprint}[1]{\href{http://ascl.net/#1}{\nolinkurl{http://ascl.net/#1}}}
\providecommand{\doarXiv}[1]{\href{https://arxiv.org/abs/#1}{\nolinkurl{https://arxiv.org/abs/#1}}}

\bibitem[{{Aartsen} {et~al.}(2013{\natexlab{a}})}]{Aartsen+13}
{Aartsen}, M.~G., {et~al.} 2013{\natexlab{a}}, \prl, 111, 021103,
  \dodoi{10.1103/PhysRevLett.111.021103}

\bibitem[{{Aartsen} {et~al.}(2013{\natexlab{b}})}]{Aartsen+13a}
---. 2013{\natexlab{b}}, \apj, 779, 132, \dodoi{10.1088/0004-637X/779/2/132}

\bibitem[{{Aartsen} {et~al.}(2014)}]{ice_3yrs}
---. 2014, \prl, 113, 101101, \dodoi{10.1103/PhysRevLett.113.101101}

\bibitem[{{Aartsen} {et~al.}(2017{\natexlab{a}})}]{Aartsen+17}
---. 2017{\natexlab{a}}, Journal of Instrumentation, 12, P03012,
  \dodoi{10.1088/1748-0221/12/03/P03012}

\bibitem[{{Aartsen} {et~al.}(2017{\natexlab{b}})}]{ice_7yr}
---. 2017{\natexlab{b}}, \apj, 835, 151, \dodoi{10.3847/1538-4357/835/2/151}

\bibitem[{{Aartsen} {et~al.}(2017{\natexlab{c}})}]{ice_stacking}
---. 2017{\natexlab{c}}, \apj, 849, 67, \dodoi{10.3847/1538-4357/aa8dfb}

\bibitem[{{Aartsen} {et~al.}(2018{\natexlab{a}})}]{txs0506a}
---. 2018{\natexlab{a}}, Science, 361, eaat1378,
  \dodoi{10.1126/science.aat1378}

\bibitem[{{Aartsen} {et~al.}(2018{\natexlab{b}})}]{txs0506b}
---. 2018{\natexlab{b}}, Science, 361, 147, \dodoi{10.1126/science.aat2890}

\bibitem[{{Aartsen} {et~al.}(2020{\natexlab{a}})}]{ice_10yrs}
---. 2020{\natexlab{a}}, \prl, 124, 051103,
  \dodoi{10.1103/PhysRevLett.124.051103}

\bibitem[{{Aartsen} {et~al.}(2020{\natexlab{b}})}]{Aartsen+20}
---. 2020{\natexlab{b}}, \apj, 898, 117, \dodoi{10.3847/1538-4357/ab9fa0}

\bibitem[{{Abbasi} {et~al.}(2021)}]{ice_track}
{Abbasi}, R., {et~al.} 2021, arXiv e-prints, arXiv:2101.09836,
  \dodoi{10.48550/arXiv.2101.09836}

\bibitem[{Abbasi {et~al.}(2021)}]{IceCube:2021mzg}
Abbasi, R., {et~al.} 2021, PoS, ICRC2021, 1073, \dodoi{10.22323/1.395.1073}

\bibitem[{{Abbasi} {et~al.}(2022)}]{ngc1068}
{Abbasi}, R., {et~al.} 2022, Science, 378, 538, \dodoi{10.1126/science.abg3395}

\bibitem[{{Abbasi} {et~al.}(2023{\natexlab{a}})}]{gal}
---. 2023{\natexlab{a}}, Science, 380, 1338, \dodoi{10.1126/science.adc9818}

\bibitem[{{Abbasi} {et~al.}(2023{\natexlab{b}})}]{Abbasi+23}
---. 2023{\natexlab{b}}, \apj, 956, 20, \dodoi{10.3847/1538-4357/acf713}

\bibitem[{{Abbasi} {et~al.}(2025)}]{Abbasi+25}
---. 2025, \apj, 981, 131, \dodoi{10.3847/1538-4357/ada94b}

\bibitem[{{Abdalla} {et~al.}(2018{\natexlab{a}})}]{hpwn18}
{Abdalla}, H., {et~al.} 2018{\natexlab{a}}, \aap, 612, A2,
  \dodoi{10.1051/0004-6361/201629377}

\bibitem[{{Abdalla} {et~al.}(2018{\natexlab{b}})}]{hgps18}
---. 2018{\natexlab{b}}, \aap, 612, A1, \dodoi{10.1051/0004-6361/201732098}

\bibitem[{{Acero} {et~al.}(2013)}]{ace+13}
{Acero}, F., {et~al.} 2013, \apj, 773, 77, \dodoi{10.1088/0004-637X/773/1/77}

\bibitem[{{Ackermann} {et~al.}(2011)}]{ack+11}
{Ackermann}, M., {et~al.} 2011, \apj, 726, 35,
  \dodoi{10.1088/0004-637X/726/1/35}

\bibitem[{{Adri{\'a}n-Mart{\'\i}nez} {et~al.}(2016)}]{km3net}
{Adri{\'a}n-Mart{\'\i}nez}, S., {et~al.} 2016, Journal of Physics G Nuclear
  Physics, 43, 084001, \dodoi{10.1088/0954-3899/43/8/084001}

\bibitem[{{Ahlers} \& {Murase}(2014)}]{Ahlers+14}
{Ahlers}, M., \& {Murase}, K. 2014, \prd, 90, 023010,
  \dodoi{10.1103/PhysRevD.90.023010}

\bibitem[{{Albert} {et~al.}(2020)}]{Albert+20}
{Albert}, A., {et~al.} 2020, \apj, 905, 76, \dodoi{10.3847/1538-4357/abc2d8}

\bibitem[{{Amato} {et~al.}(2003){Amato}, {Guetta}, \& {Blasi}}]{Amato+03}
{Amato}, E., {Guetta}, D., \& {Blasi}, P. 2003, \aap, 402, 827,
  \dodoi{10.1051/0004-6361:20030279}

\bibitem[{{Ballet} {et~al.}(2023){Ballet}, {Bruel}, {Burnett}, {Lott}, \& {The
  Fermi-LAT collaboration}}]{Ballet+23}
{Ballet}, J., {Bruel}, P., {Burnett}, T.~H., {Lott}, B., \& {The Fermi-LAT
  collaboration}. 2023, arXiv e-prints, arXiv:2307.12546,
  \dodoi{10.48550/arXiv.2307.12546}

\bibitem[{{Bednarek}(2002)}]{Bednarek02}
{Bednarek}, W. 2002, \mnras, 331, 483, \dodoi{10.1046/j.1365-8711.2002.05207.x}

\bibitem[{{Bednarek}(2003)}]{Bednarek+03}
---. 2003, \aap, 407, 1, \dodoi{10.1051/0004-6361:20030929}

\bibitem[{{Bednarek} \& {Protheroe}(1997)}]{Bednarek+97}
{Bednarek}, W., \& {Protheroe}, R.~J. 1997, \prl, 79, 2616,
  \dodoi{10.1103/PhysRevLett.79.2616}

\bibitem[{Bellenghi {et~al.}(2023)}]{Bel+24}
Bellenghi, C., {et~al.} 2023, PoS, ICRC2023, 1061, \dodoi{10.22323/1.444.1061}

\bibitem[{{Belolaptikov} {et~al.}(2022)}]{baikal}
{Belolaptikov}, I., {et~al.} 2022, in 37th International Cosmic Ray Conference,
  2, \dodoi{10.22323/1.395.002}

\bibitem[{{Braun} {et~al.}(2008){Braun}, {Dumm}, {De Palma}, {Finley}, {Karle},
  \& {Montaruli}}]{Bra+08}
{Braun}, J., {Dumm}, J., {De Palma}, F., {et~al.} 2008, Astroparticle Physics,
  29, 299, \dodoi{10.1016/j.astropartphys.2008.02.007}

\bibitem[{{Cao} {et~al.}(2021)}]{Lhaaso21}
{Cao}, Z., {et~al.} 2021, Science, 373, 425, \dodoi{10.1126/science.abg5137}

\bibitem[{{Cao} {et~al.}(2024)}]{Cao+24}
---. 2024, \apjs, 271, 25, \dodoi{10.3847/1538-4365/acfd29}

\bibitem[{{Cheng} {et~al.}(1990){Cheng}, {Cheung}, {Lau}, {Yu}, \&
  {Kwok}}]{Cheng+90}
{Cheng}, K.~S., {Cheung}, T., {Lau}, M.~M., {Yu}, K.~N., \& {Kwok}, P.~W. 1990,
  Journal of Physics G Nuclear Physics, 16, 1115,
  \dodoi{10.1088/0954-3899/16/7/022}

\bibitem[{{Di Palma} {et~al.}(2017){Di Palma}, {Guetta}, \& {Amato}}]{Di+17}
{Di Palma}, I., {Guetta}, D., \& {Amato}, E. 2017, \apj, 836, 159,
  \dodoi{10.3847/1538-4357/836/2/159}

\bibitem[{{Fang} \& {Halzen}(2024)}]{Fang+24}
{Fang}, K., \& {Halzen}, F. 2024, Journal of High Energy Astrophysics, 43, 140,
  \dodoi{10.1016/j.jheap.2024.07.001}

\bibitem[{{Fang} {et~al.}(2016){Fang}, {Kotera}, {Murase}, \&
  {Olinto}}]{Fang+16}
{Fang}, K., {Kotera}, K., {Murase}, K., \& {Olinto}, A.~V. 2016, \jcap, 2016,
  010, \dodoi{10.1088/1475-7516/2016/04/010}

\bibitem[{{Fang} {et~al.}(2012){Fang}, {Kotera}, \& {Olinto}}]{Fang+12}
{Fang}, K., {Kotera}, K., \& {Olinto}, A.~V. 2012, \apj, 750, 118,
  \dodoi{10.1088/0004-637X/750/2/118}

\bibitem[{{Ferrand} \& {Safi-Harb}(2012)}]{snrcat}
{Ferrand}, G., \& {Safi-Harb}, S. 2012, Advances in Space Research, 49, 1313,
  \dodoi{10.1016/j.asr.2012.02.004}

\bibitem[{{Fujita} {et~al.}(2009){Fujita}, {Ohira}, {Tanaka}, \&
  {Takahara}}]{Fujita+09}
{Fujita}, Y., {Ohira}, Y., {Tanaka}, S.~J., \& {Takahara}, F. 2009, \apjl, 707,
  L179, \dodoi{10.1088/0004-637X/707/2/L179}

\bibitem[{{Gabici} \& {Aharonian}(2007)}]{Gabici+07}
{Gabici}, S., \& {Aharonian}, F.~A. 2007, \apjl, 665, L131,
  \dodoi{10.1086/521047}

\bibitem[{{Gabici} {et~al.}(2009){Gabici}, {Aharonian}, \&
  {Casanova}}]{Gabici+09}
{Gabici}, S., {Aharonian}, F.~A., \& {Casanova}, S. 2009, \mnras, 396, 1629,
  \dodoi{10.1111/j.1365-2966.2009.14832.x}

\bibitem[{{Guetta} \& {Amato}(2003)}]{Guetta+03}
{Guetta}, D., \& {Amato}, E. 2003, Astroparticle Physics, 19, 403,
  \dodoi{10.1016/S0927-6505(02)00221-9}

\bibitem[{{Helfand}(1979)}]{Helfand79}
{Helfand}, D.~J. 1979, \nat, 278, 720, \dodoi{10.1038/278720a0}

\bibitem[{{Ji} {et~al.}(2025){Ji}, {Wang}, \& {Zheng}}]{Ji+25}
{Ji}, S., {Wang}, Z., \& {Zheng}, D. 2025, \apj, 979, 1,
  \dodoi{10.3847/1538-4357/ad9f40}

\bibitem[{{Jiang} {et~al.}(2023){Jiang}, {Zhou}, {Zhu}, {Wang}, \&
  {Wang}}]{jiang+23}
{Jiang}, N., {Zhou}, Z., {Zhu}, J., {Wang}, Y., \& {Wang}, T. 2023, \apjl, 953,
  L12, \dodoi{10.3847/2041-8213/acebe3}

\bibitem[{{Jiang} {et~al.}(2007){Jiang}, {Chen}, \& {Zhang}}]{Jiang+07}
{Jiang}, Z.~J., {Chen}, S.~B., \& {Zhang}, L. 2007, \apj, 667, 1059,
  \dodoi{10.1086/521183}

\bibitem[{{Lemoine} {et~al.}(2015){Lemoine}, {Kotera}, \&
  {P{\'e}tri}}]{Lemoine+15}
{Lemoine}, M., {Kotera}, K., \& {P{\'e}tri}, J. 2015, \jcap, 2015, 016,
  \dodoi{10.1088/1475-7516/2015/07/016}

\bibitem[{{Li} {et~al.}(2024){Li}, {Yuan}, {He}, {Wang}, {Zhu}, {Liang},
  {Jiang}, \& {Wei}}]{Li+24}
{Li}, R.-L., {Yuan}, C., {He}, H.-N., {et~al.} 2024, arXiv e-prints,
  arXiv:2411.06440, \dodoi{10.48550/arXiv.2411.06440}

\bibitem[{{Li} {et~al.}(2025){Li}, {Huang}, {Xu}, \& {He}}]{Li+25}
{Li}, W., {Huang}, T.-Q., {Xu}, D., \& {He}, H. 2025, \apj, 980, 164,
  \dodoi{10.3847/1538-4357/adabe8}

\bibitem[{{Link} \& {Burgio}(2005)}]{Link+05}
{Link}, B., \& {Burgio}, F. 2005, \prl, 94, 181101,
  \dodoi{10.1103/PhysRevLett.94.181101}

\bibitem[{{Link} \& {Burgio}(2006)}]{Link+06}
---. 2006, \mnras, 371, 375, \dodoi{10.1111/j.1365-2966.2006.10665.x}

\bibitem[{{Matheson} {et~al.}(2016){Matheson}, {Safi-Harb}, \&
  {Kothes}}]{Matheson+16}
{Matheson}, H., {Safi-Harb}, S., \& {Kothes}, R. 2016, \apj, 825, 134,
  \dodoi{10.3847/0004-637X/825/2/134}

\bibitem[{{Murase} {et~al.}(2016){Murase}, {Guetta}, \& {Ahlers}}]{Murase+16}
{Murase}, K., {Guetta}, D., \& {Ahlers}, M. 2016, \prl, 116, 071101,
  \dodoi{10.1103/PhysRevLett.116.071101}

\bibitem[{{Murase} {et~al.}(2020){Murase}, {Kimura}, \&
  {M{\'e}sz{\'a}ros}}]{Murase+20}
{Murase}, K., {Kimura}, S.~S., \& {M{\'e}sz{\'a}ros}, P. 2020, \prl, 125,
  011101, \dodoi{10.1103/PhysRevLett.125.011101}

\bibitem[{{Neronov} {et~al.}(2024){Neronov}, {Savchenko}, \&
  {Semikoz}}]{Ner+24}
{Neronov}, A., {Savchenko}, D., \& {Semikoz}, D.~V. 2024, \prl, 132, 101002,
  \dodoi{10.1103/PhysRevLett.132.101002}

\bibitem[{{Reusch} {et~al.}(2022)}]{Reusch+22}
{Reusch}, S., {et~al.} 2022, \prl, 128, 221101,
  \dodoi{10.1103/PhysRevLett.128.221101}

\bibitem[{{Smith} {et~al.}(2023)}]{Smith+23}
{Smith}, D.~A., {et~al.} 2023, \apj, 958, 191, \dodoi{10.3847/1538-4357/acee67}

\bibitem[{{Stein} {et~al.}(2021)}]{Stein+21}
{Stein}, R., {et~al.} 2021, Nature Astronomy, 5, 510,
  \dodoi{10.1038/s41550-020-01295-8}

\bibitem[{{Straal} \& {van Leeuwen}(2019)}]{Straal+19}
{Straal}, S.~M., \& {van Leeuwen}, J. 2019, \aap, 623, A90,
  \dodoi{10.1051/0004-6361/201833922}

\bibitem[{{Taylor} {et~al.}(1996){Taylor}, {Goss}, {Coleman}, {van Leeuwen}, \&
  {Wallace}}]{Taylor+96}
{Taylor}, A.~R., {Goss}, W.~M., {Coleman}, P.~H., {van Leeuwen}, J., \&
  {Wallace}, B.~J. 1996, \apjs, 107, 239, \dodoi{10.1086/192363}

\bibitem[{{Taylor} {et~al.}(1992){Taylor}, {Wallace}, \& {Goss}}]{Taylor+92}
{Taylor}, A.~R., {Wallace}, B.~J., \& {Goss}, W.~M. 1992, \aj, 103, 931,
  \dodoi{10.1086/116115}

\bibitem[{{van Velzen} {et~al.}(2024)}]{van+24}
{van Velzen}, S., {et~al.} 2024, \mnras, 529, 2559,
  \dodoi{10.1093/mnras/stae610}

\bibitem[{{Vecchiotti} {et~al.}(2023){Vecchiotti}, {Villante}, \&
  {Pagliaroli}}]{vec+23}
{Vecchiotti}, V., {Villante}, F.~L., \& {Pagliaroli}, G. 2023, \apjl, 956, L44,
  \dodoi{10.3847/2041-8213/acff60}

\bibitem[{{Wallace} {et~al.}(1997){Wallace}, {Landecker}, \&
  {Taylor}}]{Wallace+97}
{Wallace}, B.~J., {Landecker}, T.~L., \& {Taylor}, A.~R. 1997, \aj, 114, 2068,
  \dodoi{10.1086/118627}

\bibitem[{Wilks(1938)}]{Wilks}
Wilks, S.~S. 1938, Annals Math. Statist., 9, 60,
  \dodoi{10.1214/aoms/1177732360}

\bibitem[{Wolf(2021)}]{Wolf+19}
Wolf, M. 2021, PoS, ICRC2019, 1035, \dodoi{10.22323/1.358.1035}

\bibitem[{{Yuan} {et~al.}(2024){Yuan}, {Winter}, \& {Lunardini}}]{yuan+24}
{Yuan}, C., {Winter}, W., \& {Lunardini}, C. 2024, \apj, 969, 136,
  \dodoi{10.3847/1538-4357/ad50a9}

\bibitem[{{Zhou} {et~al.}(2023)}]{Zhou+23}
{Zhou}, X., {et~al.} 2023, \apjs, 268, 61, \dodoi{10.3847/1538-4365/acee7f}

\end{thebibliography}

\appendix
\restartappendixnumbering

\section{LAT Data analysis for 4FGL J1947.7+2744}
\label{sec:fermiana}
\begin{figure*}
\centering
\includegraphics[width=0.54\textwidth]{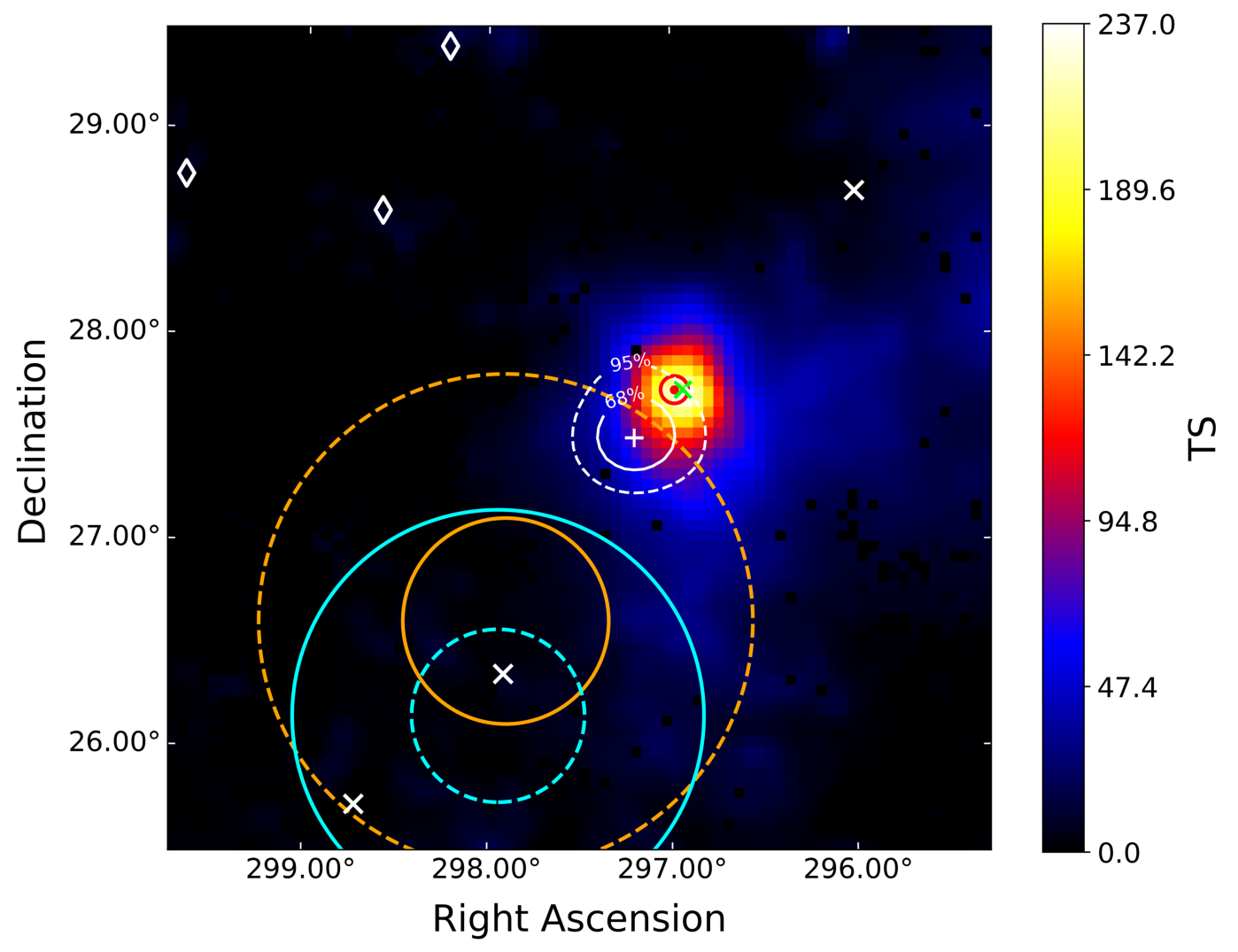}
\includegraphics[width=0.43\textwidth]{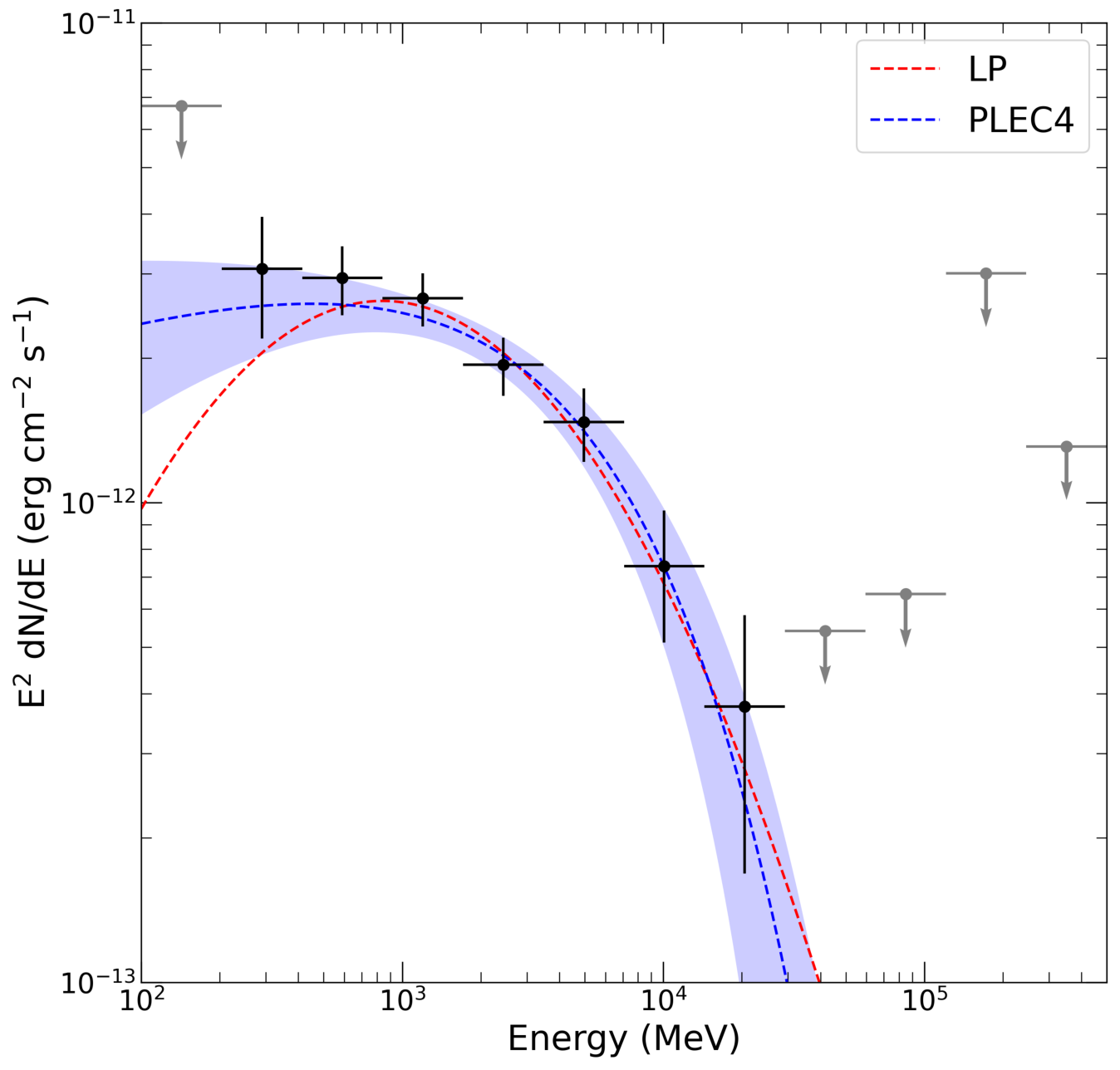}
	\caption{{\it Left:} \fermi\ LAT TS map of the G63.7+1.1 region in
	0.1--500\,GeV. The white diamonds and crosses mark 
	the identified/associated and
	unassociated \fermi\ \gr\ sources, respectively, in the field. 
	Other marked are the same as those in Figure~\ref{fig:tsmap}. 
	{\it Right:} $\gamma$-ray spectrum of 4FGL~J1947.7+2744. Both 
	the best-fit LP and PLEC4 model spectra (red dashed and blue dashed 
	respectively) are shown.
\label{fig:fermi_tsmap}}
\end{figure*}
We selected photon events in the energy range of 0.1--500 GeV 
(evclass=128 and evtype=3) from the updated Fermi Pass 8 database in a time 
range of from 2008-08-04 15:43:36 (UTC) to 2024-04-18 00:05:53 (UTC).
The region of interest was 20$^\circ$ $\times$ 20$^\circ$, centered
at the position of 4FGL J1947.7+2744. 
Events with zenith angles $>$90$^\circ$\ were excluded to avoid 
the contamination from the Earth limb, and the expression 
DATA\_QUAL $>$ 0 \&\& LAT\_CONFIG = 1 was used for selecting good time-interval 
events. The package Fermitools-2.2.0 and the instrumental 
response function P8R3\_SOURCE\_V3 were used in our analysis.

The source model was constructed from 4FGL-DR4. All sources 
in the catalog within 25$^\circ$\ of the target were included. The spectral 
models in 4FGL-DR4 for the sources were adopted.
Spectral indices and normalizations of the sources within 
5$^\circ$\ of the target were set as free parameters and all the other 
parameters were fixed
at the catalog values. The extragalactic diffuse emission 
and the Galactic diffuse emission components, the spectral files 
iso\_P8R3\_SOURCE\_V3\_v1.txt and gll\_iem\_v07.fits respectively, were 
included.
The normalizations of these two components were always set as free parameters 
in our analysis.

In 4FGL-DR4, 4FGL J1947.7+2744 was modeled as a point source with a 
Log-Parabola (LP) spectrum,
$dN/dE = N_0 (E/E_b)^{-[\alpha +\beta\ln(E/E_b)]}$, where $E_b$ was fixed 
at 1.9\,GeV. Setting this LP spectral model, we performed the standard binned 
likelihood analysis to the whole data in 0.3--500\,GeV. We obtained 
its flux $F_{\gamma} = (4.72\pm0.62)\times$10$^{-9}$~ph\,cm$^{-2}$\,s$^{-1}$, 
with a TS value of 257 (detection significance 
$\approx$$\sqrt{{\rm TS}} \approx 16\sigma$), while the other
obtained spectral parameters are given in Table~\ref{tab:para} of this section.
The results are consistent with those in 4FGL-DR4. As we suspected that
the $\gamma$-ray emission could likely be that of a pulsar, we also used  
the typical pulsar model PLEC4 ({\tt PLSuperExpCutoff4} in Fermitools\footnote{\url{ https://fermi.gsfc.nasa.gov/ssc/data/analysis/software/.}}), 
$dN/dE = N_0 (E/E_0)^{-\Gamma + d/b} e^{d/b^2[1-(E/E_0)^b]}$, 
where $E_0$ and $b$ were fixed at the typical values of 1.6\,GeV and 2/3, 
respectively \citep{Smith+23}.
We obtained  
$F_{\gamma} = (4.93\pm0.61)\times$10$^{-9}$~ph\,cm$^{-2}$\,s$^{-1}$, with
a TS value of 257, 
and the values of the other parameters are also given in Table \ref{tab:para}. 
A TS map of the source region is shown in the left panel of 
Figure~\ref{fig:fermi_tsmap} of this section.
The same TS value, and also the same likelihood value in calculating
the TS value as we checked, 
from PLEC4 indicates that it provides a fit as well as LP.
We adopted PLEC4 in the following analysis.

We tested the extension of 4FGL J1947.7+2744 using the $\gamma$-ray data above 
1\,GeV. The spatial template of a uniform disk with radius from 
0$^\circ.$1--1$^\circ.$0 (at a step of 0$^\circ.$1) was used. We did not 
detect any extension for the source. 

We obtained the spectral data points for 4FGL~J1947.7+2744 by performing 
the binned likelihood analysis to the data in 12 energy bins 
evenly divided in logarithm from 0.1 to 500\,GeV. In the analysis, the 
spectral normalizations of the sources in the source model within 5$^\circ$\ of 
the target were set free and all other spectral parameters of 
the sources were fixed at the values obtained in the above likelihood analysis.
For the obtained results, the data points with TS $\geq$ 4 were kept and 
otherwise the 95\% upper limits were calculated. The $\gamma$-ray spectral data
points and the upper limits are shown in the right panel of 
Figure \ref{fig:fermi_tsmap}.

\begin{table}
    \centering
    \caption{LAT likelihood analysis results}
    \begin{tabular}{lccc}
        \hline
        \hline
        Model &  & Parameter \\
        \hline
        LP & $\alpha$ & $\beta$ & TS   \\
        Catalog & 2.36$\pm$0.10 & 0.22$\pm$0.07 & 230  \\
        This work & 2.34$\pm$0.09 & 0.22$\pm$0.07 & 257 \\
        \hline
        PLEC4 & $\Gamma$ & $d$ &  TS  \\
             & 2.23$\pm$0.10 & 0.27$\pm$0.10 & 257 \\
        \hline
    \end{tabular}
    \label{tab:para}
\end{table}

\section{4FGL J1947.7+2744 as the pulsar}
\label{sec:psr}

\begin{table}
    \centering
	\caption{Luminosity comparison (with distance at 6\,kpc)}
    \begin{tabular}{lcccc}
        \hline
        \hline
        ~ & PWN & Pulsar candidate \\
        \hline
	    $L_{X}$ (erg~s$^{-1}$) & $1.6\times10^{33}$ & $1.1\times10^{32}$ \\\hline
        $L_{\gamma}$ (erg~s$^{-1}$) & ~ ~ ~ ~ ~ ~ ~ ~ ~ ~ $3.2\times10^{34}$\\
        $L_{\gamma}$/$L_{X}$ & 20 & 290 \\\hline
	    $L_{\rm TeV}$ (erg~s$^{-1}$) & $\leq 3.2\times10^{33}$ &   \\
        $L_{\gamma}$/$L_{\rm TeV}$ & $\geq$10 &  \\
        \hline
    \end{tabular}
    \tablecomments{$L_{X}$ values are from \citealt{Matheson+16}. $L_{\rm TeV}$ is the 95\% upper limit in 1--10\,TeV from the HESS Galactic plane survey.}
    \label{tab:lum}
\end{table}

No radio pulsar in G63.7+1.1 has been reported \citep{Straal+19}.
A pulsar candidate was found at the X-ray intensity peak of G63.7+1.1 by \citealt{Matheson+16}, and on the basis of the X-ray properties, they 
estimated the spin-down energy $\dot{E}$ and the characteristic age 
$\tau_{\rm c}$ of the putative pulsar to be $2.1 \times 10^{36}$\,erg\,s$^{-1}$ 
and 27\,kyr, respectively, where the distance $D$ was considered to 
be $\sim$6\,kpc.
The \fermi\ LAT source 4FGL J1947.7+2744 is listed as a PWN in 4FGL-DR4. 
However, we argue that it is more likely the emission of the pulsar. 
First, only a few 
PWNe have been reported to have significant GeV $\gamma$-ray emission \citep{ack+11, ace+13}
, while on the other hand, for example, there 
are $\gtrsim$30 PWNe being detected at TeV energies with the High Energy 
Stereoscopic System (HESS; \citealt{hpwn18})
and other facilities \citep{Aartsen+20}.  PWNe are
more likely to be detected at TeV $\gamma$-rays. According to \citealt{ace+13}, the GeV-to-TeV luminosity ratio $L_{\gamma}/L_{\rm TeV}$ 
for PWNe is around $\sim$2.7.
In Table~\ref{tab:lum}, we listed the X-ray luminosities 
of the PWN G63.7+1.1 and the candidate pulsar and the 95\% upper limit on
the 1--10\,TeV luminosity $L_{\rm TeV}$ of the PWN.
The TeV upper limit was obtained from the HESS survey data \citep{hgps18} by 
assuming a 0$^\circ.$1 circular region
at the PWN's X-ray position (the typical point-spread function of 
the HESS imaging had a size of 0$^\circ.$08 with $\pm$20\% variations)
and a power-law (PL) emission with 
photon index $\Gamma =$ 2.4 (a typical value for TeV PWNe; 
see \citealt{hpwn18}). 
If we assume 4FGL J1947.7+2744 as the GeV PWN,
$L_{\gamma}/L_{\rm TeV} \geq 10$, which is greater than those of
all other PWNe. Secondly, the GeV emission can be well described
with the typical pulsar model PLEC4, and the corresponding
$\gamma$-ray luminosity 
$L_\gamma \simeq 3.2 \times 10^{34} D_6^2$\,erg\,s$^{-1}$ (where
$D_6$ is $D$ scaled by 6\,kpc) and $\gamma$-ray efficiency 
$\eta = L_\gamma / \dot{E} \simeq 1.5\times 10^{-2} D_6^{0.62}$ 
($\dot{E}\sim D^{1.38}$; see \citealt{Matheson+16}) are in the ranges
of \gr\ pulsars for the estimated $\dot{E}$ and $\tau_{\rm c}$ \citep{Smith+23}.
In addition, considering this putative pulsar as a radio-quiet (RQ) one, its
$L_{\gamma}/L_X\sim 290$, also in the range of RQ \gr\ pulsars (10$^2$--10$^4$)
when $\tau_{\rm c} \gtrsim 10$\,kyr (J. Zheng et al., in preparation). The
radio-loud \gr\ pulsars have the ratio in a wider range, with the low-end value
being $\sim$10. Thus, the GeV \gr\ source is likely the pulsar instead, which
may be verified by a deep radio search.


\section{Background TS distribution}
\label{sec:back}

\begin{figure*}
\centering
\includegraphics[width=0.5\textwidth]{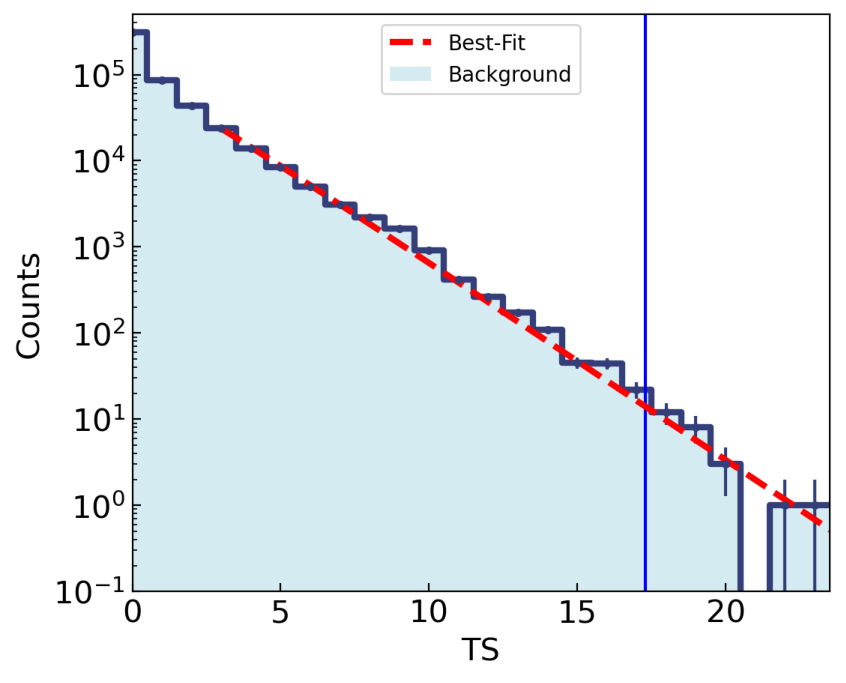}
	\caption{Background TS distribution (blue bars) and best-fit model 
	using a truncated gamma function (dashed red line), which are used to 
	estimate the p-value. The vertical line shows the TS value at the 
	location of G63.7+1.1.
\label{fig:back}}
\end{figure*}

In Figure~\ref{fig:back}, the background TS distribution at the location of
the PWN G63.7+1.1 is shown.

\end{document}